\documentclass{appolb}
\usepackage{epsfig}
\newcommand{\ta}[1]{#1\hspace{-.42em}/\hspace{-.07em}} % SLASH
\newcommand{\taa}[1]{#1\hspace{-.55em}/\hspace{-.09em}} % SLASH
\newcommand{\beq}{\begin{equation}} 
\newcommand{\eeq}{\end{equation}}   
\newcommand{\bea}{\begin{eqnarray}} 
\newcommand{\eea}{\end{eqnarray}}

\begin{document}
\date{\today}
\pagestyle{plain}
%% uncomment the following line to get equations numbered by (sec.num)
%\eqsec
\newcount\eLiNe\eLiNe=\inputlineno\advance\eLiNe by -1
\title{ Radiative return method as a tool in hadronic physics
\thanks{Presented by H. Czy{\.z}
 at XXIX International Conference of Theoretical Physics,
 `Matter To The Deepest', Ustro{\'n}, 8-14 September 2005, Poland.
 Work supported in part by
 EC 5-th Framework Program under contract
  HPRN-CT-2002-00311 (EURIDICE network),  TARI project RII3-CT-2004-506078
 and
  Polish State Committee for Scientific Research
  (KBN) under contracts 1 P03B 003 28 and 1 P03B 013 27.}
}
\author{Henryk Czy\.z$^a$
, Agnieszka Grzeli\'nska$^b$\\ and El\.zbieta~Nowak-Kubat$^a$
\address{ a: \ Institute of Physics, University of Silesia,
PL-40007 Katowice, Poland \\ b: \ Institut f\"ur Theoretische Teilchenphysik,
Universit\"at Karlsruhe,\\ D-76128 Karlsruhe, Germany 
 }}
\maketitle

 \vskip -5 cm
 \hfill TTP05-20
 \vskip 5 cm

\begin{abstract}
 A short review of both theoretical and experimental aspects 
 of the radiative return method is presented. It is emphasised
 that the method gives not only possibility of the independent
 from the scan method measurement of the hadronic cross section,
 but also can provide information concerning details of the hadronic
 interactions. New developments in the PHOKHARA event generator
 are also reviewed. The 3 pion and kaon pair production is implemented
 within the version 5.0 of the program, together with contributions
 of the radiative $\phi$ decays to the 2 pion final states.
 Missing NLO radiative corrections to the $e^+e^- \to \mu^+\mu^- \gamma$
 process will be implemented in the forthcoming version of the generator.
 \end{abstract}

%***********************************************************************
\section{Introduction}
%***********************************************************************
 Precise hadronic cross section measurements are crucial
 for predictions of the hadronic contributions to $a_\mu$, the anomalous 
 magnetic moment of the muon, and to the running of the electromagnetic
 coupling ($\alpha_{QED}$) from its value at low energy up
 to $M_Z$ (for recent reviews look
 \cite{Davier:2003pw,Jegerlehner:2003rx,Nyffeler:2004mw}).
 When using the scan method one usually needs new experiments to
 be performed with not negligible costs, while
 the radiative return method proposed already years ago \cite{Zerwas}
 allows for extracting the information on the hadronic cross section
 basing on the measurements at the existing meson factories, profiting
 from their huge luminosities and excellent detectors.
 As it was shown in  \cite{Nowak,Czyz:2004nq} the radiative
 return method 
  is not only a powerful tool in the 
 measurement of $\sigma (e^+e^- \to \  {\mathrm{hadrons}})$,
 but allows for detailed studies of hadronic interactions.
  Due to a complicated experimental setup,
 the use of Monte Carlo (MC) event generators 
 \cite{Szopa,rest,FSR,PHOKHARA_mu,actaHN,Czyz:2005},
 which include various radiative corrections \cite{PHradcor}
 is indispensable for both signal and background processes.
 Some more detailed analysis of that subject can be found
 also in \cite{proc,FSR1}.
 
 This paper is aimed as a short review of the results obtained 
 by the radiative return method. It presents also a further
 potential of the method and shows new developments in the
 PHOKHARA event generator important for an extraction
 of the $\sigma(e^+e^-\to {\rm hadrons})$ (and other physical 
 quantities) from the measured cross section 
 $\sigma(e^+e^-\to {\rm hadrons} + {\rm photons})$.

 %*******************************************************************
\section{Hadronic cross section \label{sec2}}

 The extraction of the cross section $\sigma(e^+e^-\to {\rm hadrons})$
 from the measured cross section 
 $\sigma(e^+e^-\to {\rm hadrons} + {\rm photons})$ relies on the
 factorisation
\bea
{ d\sigma(e^+e^- \to \mathrm{hadrons} + n \gamma) = }
{\ H }{ \ d\sigma(e^+e^-\to \mathrm{hadrons})}
\label{fac}
\eea
valid at any order for photons emitted from initial leptons, where
 the function $H$ contains QED radiative corrections. It is known
 analytically, if no cuts are imposed on photons, at next to leading order
 (NLO) and has to be provided
 in form of an event generator of the reaction  
 $e^+e^-\to {\rm hadrons} + {\rm photons}$ 
 for a realistic experimental setup. The emission of photons from
 the final state hadrons has to be controlled as well \cite{FSR,FSR1},
  with an accuracy
 which allows for an error small enough not to spoil the accuracy 
 of the  $\sigma(e^+e^-\to {\rm hadrons})$ extraction.

 Comparing the
  scan method and the radiative return method one has to say that
 they are in many aspects complementary. It is due to the fact that
 many experimental systematic errors are completely different in both cases.
 The radiative return method has though the advantage that most of them are
 the same for all the values of the invariant mass of the hadronic system
  for which the measurement is performed. That is not true for the scan method,
 where for each energy one has to perform a separate analysis of many
 systematical errors (an energy calibration etc.).  It is also important
 that using the radiative return method one can use machines, which were
 built for other purposes ($\phi$ and B factories) and 
 one has to `invest' only in the experimental analysis. In many aspects
 both methods encounter  however the same problems, which have to be solved.
 The already mentioned
 final state emission has to be studied carefully in both cases, 
 the photon vacuum polarisation with its fast varying behaviour
 nearby resonances has to be taken into account,
  higher orders radiative corrections have to be properly implemented
 in event generators etc. Even if
 the details are different for scan and the radiative return method
 the main features remain the same.

 The most
 important hadronic channel,
  from the point of the hadronic contributions to $a_\mu$,
  mainly $\pi^+\pi^-$, is an example of that complementarity.
 Very accurate KLOE measurement \cite{KLOE1} provided an important
 cross check of the CMD-2 data \cite{Akhmetshin:2003zn}
  and even if the agreement 
 is not excellent it has allowed to conclude that the disagreement
 between $e^+e^-$ data and the $\tau$ data concerning the pion form factor
 is not of the experimental origin. Further, mostly theoretical work,
 will be required to solve that puzzle and one will have to
 find new sources of the 
 isospin violation effects, which finally will explain that disagreement.

 Already now many new valuable physical information was obtained by means
 of the radiative return method. The BaBar measurement of the 
 $\sigma(e^+e^- \to 2\pi^+2\pi^-,2K^+2K^-,K^+K^-\pi^+\pi^-)$
  \cite{Aubert:2005eg}
 are the most accurate to date
 results, with the $2\pi^+2\pi^-$ mode also extremely important 
 for hadronic contributions to $a_\mu$ and $\alpha_{QED}$.
 The BaBar measurement of the $\sigma(e^+e^- \to \pi^+\pi^-\pi^0)$
  \cite{Aubert:2004kj}
  has shown that the cross section 
 around the $\omega$'' resonance is actually much bigger as compared
  to an
   old measurement by DM2 collaboration \cite{Antonelli:92}.
 Results coming from BaBar on narrow resonances 
\cite{Aubert:2003sv,Aubert:2004kj,Aubert:2005eg}
 are also very competitive to the ones coming from the scan method
 (for a review see \cite{Brambilla:2004wf}).
  Not to mention the forthcoming results \cite{anulli,Marianna}, which
 show still growing potential for the radiative return method.

%*******************************************************************
\section{Not only the hadronic cross section - looking inside the hadronic
 interactions}

The radiative return method originally proposed for the hadronic cross
 section measurements \cite{Zerwas,Binner} can be used to much more
 detailed studies of the hadronic interaction. The first investigations
 along these lines were done in \cite{Nowak}, where it was shown that
 it is feasible to measure separately the nucleon form factors in the
 time-like region at B-factories. That measurements are important for
 the understanding of the experimental situation 
  in the space-like region (for a review look \cite{arrington}), where
 two different type of measurements lead to different results for
 the ratio of the magnetic and electric proton form factors.

 The nucleon electromagnetic current is defined by the 
 form factors as follows

\bea
 J_\mu =  - i e \cdot \bar u(q_2)
\left({ F_1^N(Q^2)}\gamma_\mu
 - \frac{{ F_2^N(Q^2)}}{4 m_N} \left[\gamma_\mu,\taa Q \ \right]
 \right) v(q_1) \ , \
\label{ff}
\eea
with electric and magnetic form factors
 $G_M^N = F_1^N + F_2^N~, \ G_E^N = F_1^N + \tau F_2^N~$.

\begin{figure}[ht]
\begin{center}
\epsfig{file=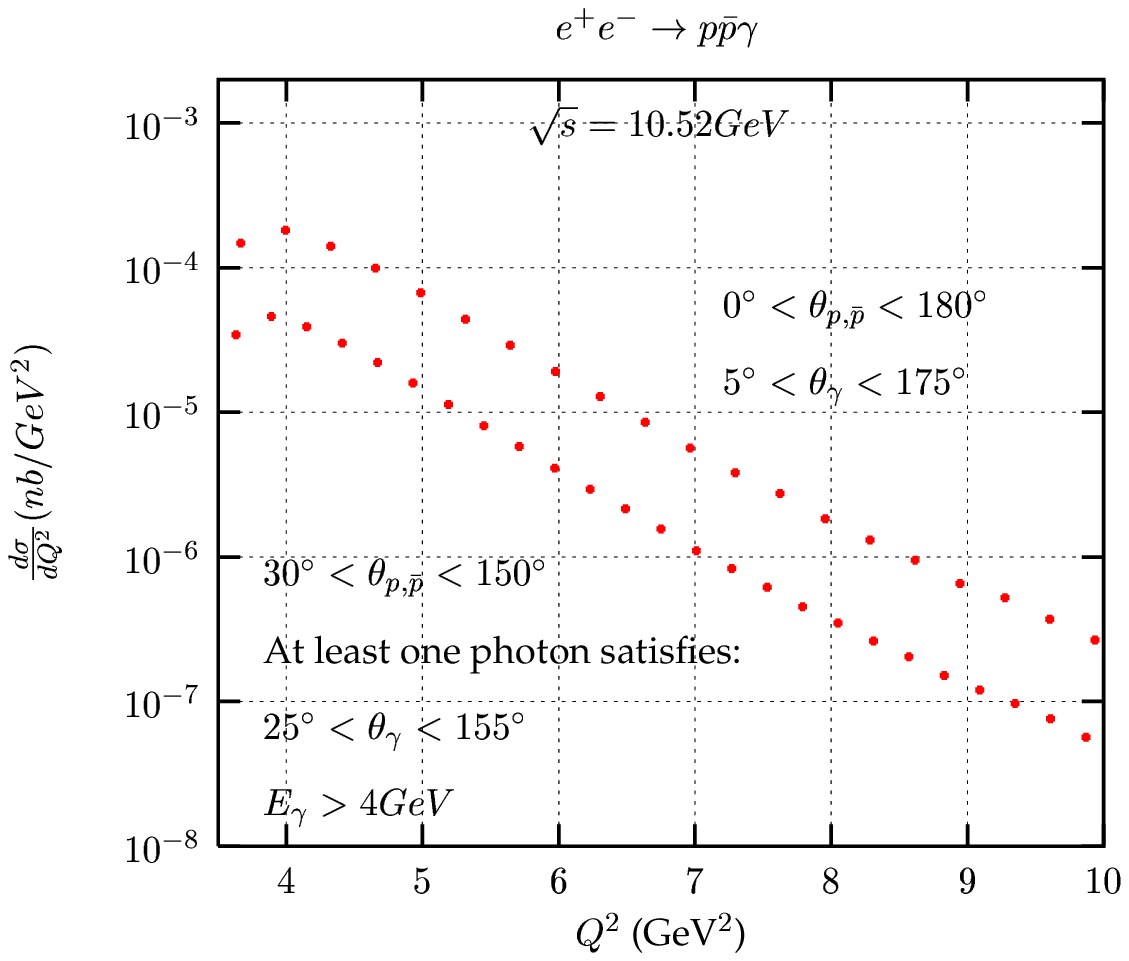,width=6.09cm} 
\epsfig{file=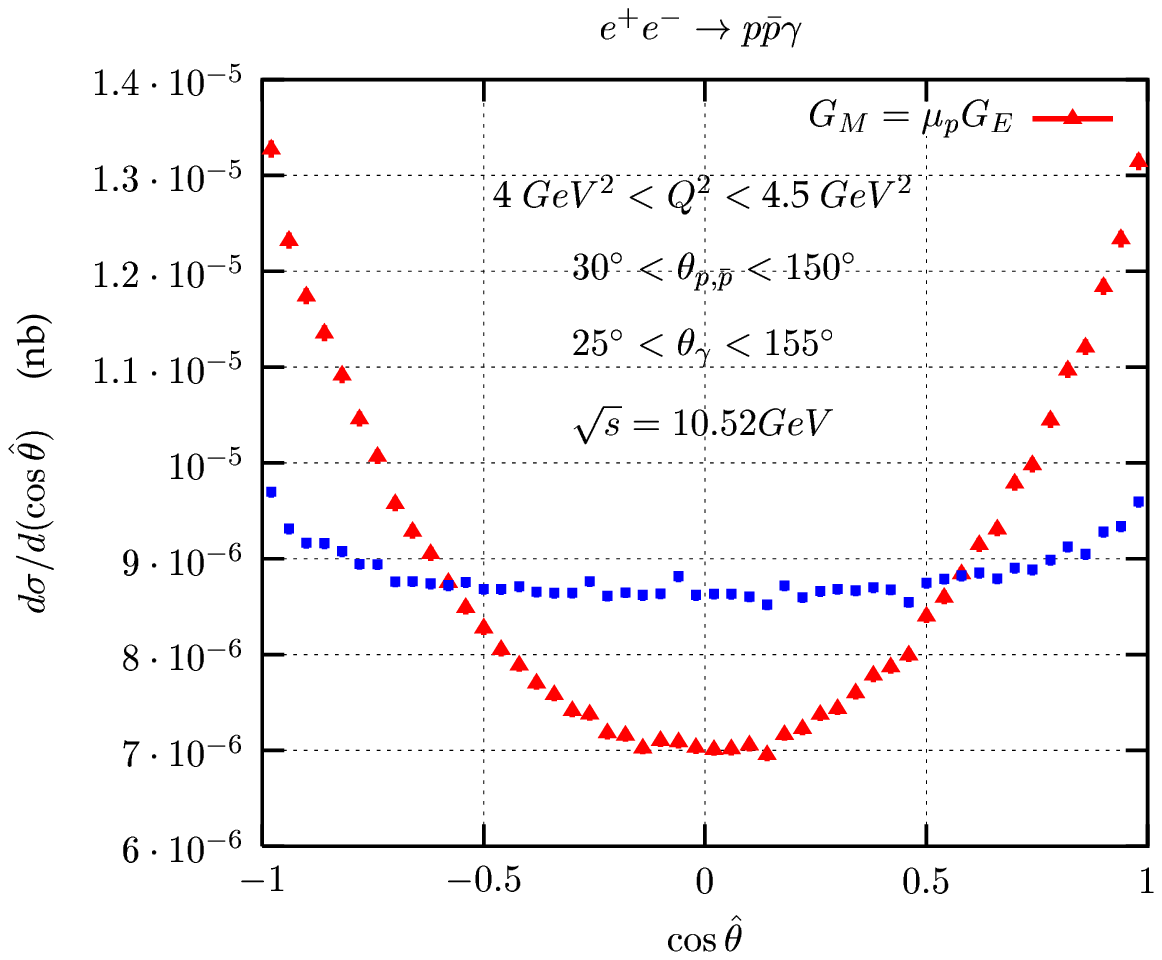,width=6.29cm} \\
 \hskip 1.5 cm (a) \hskip 5.6 cm (b)
\caption{ (a) The differential, in $Q^2$, cross section for the reaction 
 $e^+e^- \to  p \bar {p} \gamma$ for two different sets of event selections.
 (b) The differential, in $\cos\hat\theta$ 
 ($\hat\theta$ - proton polar angle in the proton pair rest frame),
  cross section for the reaction 
 $e^+e^-\to \bar p p \gamma$ for 
 two theoretical models (see text for details).}
\label{nucleon}
\end{center}
\end{figure}

 The statistics is not a problem
 for a measurement at B-factories with hundreds of fb$^{-1}$ accumulated
 luminosity as the integrated cross section
 for the event selection corresponding
 to lower curve in 
 Fig.~\ref{nucleon}a (angular cuts close to BaBar angular acceptance)
 is about 59.3~fb for protons and 125~fb for neutrons
 in the final state.  The separation of the electric and magnetic form factors
 is also possible for quite a big range of the nucleon pair
 invariant mass. It is particularly easy if one performs analysis 
 in the nucleon
 pair rest frame as shown in Fig.~\ref{nucleon}b, where the proton
 polar angle distribution is plotted both for a model which predict
 the ratio of form factors in agreement with measurements using the
 Rosenbluth method ($G_M^p = \mu_p G_E^p$, triangles) and a model which
 predict the ratio of the form factors with agreement with the measurements
 using polarisation method (squares).
  For details concerning both methods see a review
 article \cite{arrington} and references therein.
 For the 
 $4 \ {\rm GeV^2} < Q^2 < 4.5 \ {\rm GeV^2}$ one expects about 2000 events
 per  100~fb$^{-1}$ accumulated luminosity and clearly two parameter 
 ($G_M^N,G_E^N$) fit is possible and its accuracy will be limited mostly
 by systematic errors.  

 Already now BaBar collaboration has preliminary results
 for the proton electromagnetic form factor measurements \cite{anulli}, 
 which when completed will allow for extensive tests of the theoretical models.

\begin{figure}[ht]
\begin{center}
\epsfig{file=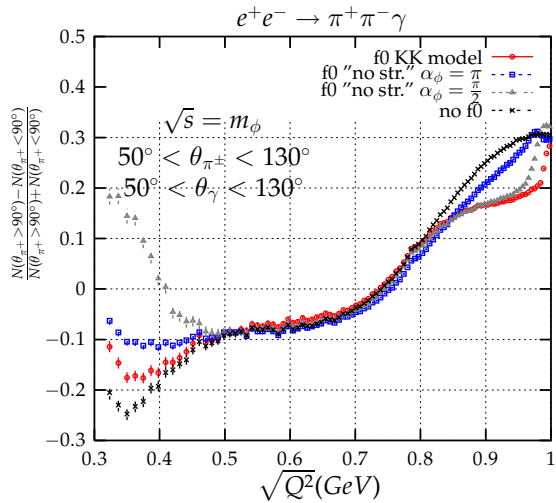,width=8cm} 
\caption{ The pion charge asymmetry for different 
radiative $\phi$ decay models \cite{Czyz:2004nq}.}
\label{asym}
\end{center}
\end{figure}

 Another example \cite{Czyz:2004nq}, very specific for the 
radiative return method
 at DAPH\-NE energy,
 is the study of radiative $\phi$ decays at KLOE (for present status of
 the experimental situation see \cite{Marianna}). The $\phi$ decay 
 ($\phi\to f_0(\to\pi^+\pi^-)\gamma$) contributing to the reaction
 $e^+e^-\to\pi^+\pi^-\gamma$ might in principle cause some problems
 in the pion form factor extraction from the 
$\sigma(e^+e^-\to\pi^+\pi^-\gamma)$ measurement. However, as it was shown
 in  \cite{Czyz:2004nq}, the detailed studies of the charge asymmetries
 allow not only to control that contribution, but also to distinguish
 between different models of the radiative $\phi$ decays. That is clearly seen
 in Fig.~\ref{asym}, where the charge asymmetry for an event selection
 enhancing the FSR contributions is presented. The differences between
 predictions coming from different models of the radiative $\phi$ decays
 and also from scalar QED (see \cite{Czyz:2004nq} for details) are
 sizable in the region of large and small values of the pion pair
 invariant mass ($Q^2$) and a measurement can easily distinguish
 between them leading to tests with unprecedented accuracy.  

%

%*******************************************************************
%*******************************************************************
\section{New developments in the PHOKHARA event generator}
%*******************************************************************
\subsection{PHOKHARA 5.0: $\pi^+\pi^-\pi^0$ and $KK$ final states}
\begin{figure}[ht]
\begin{center}
\epsfig{file=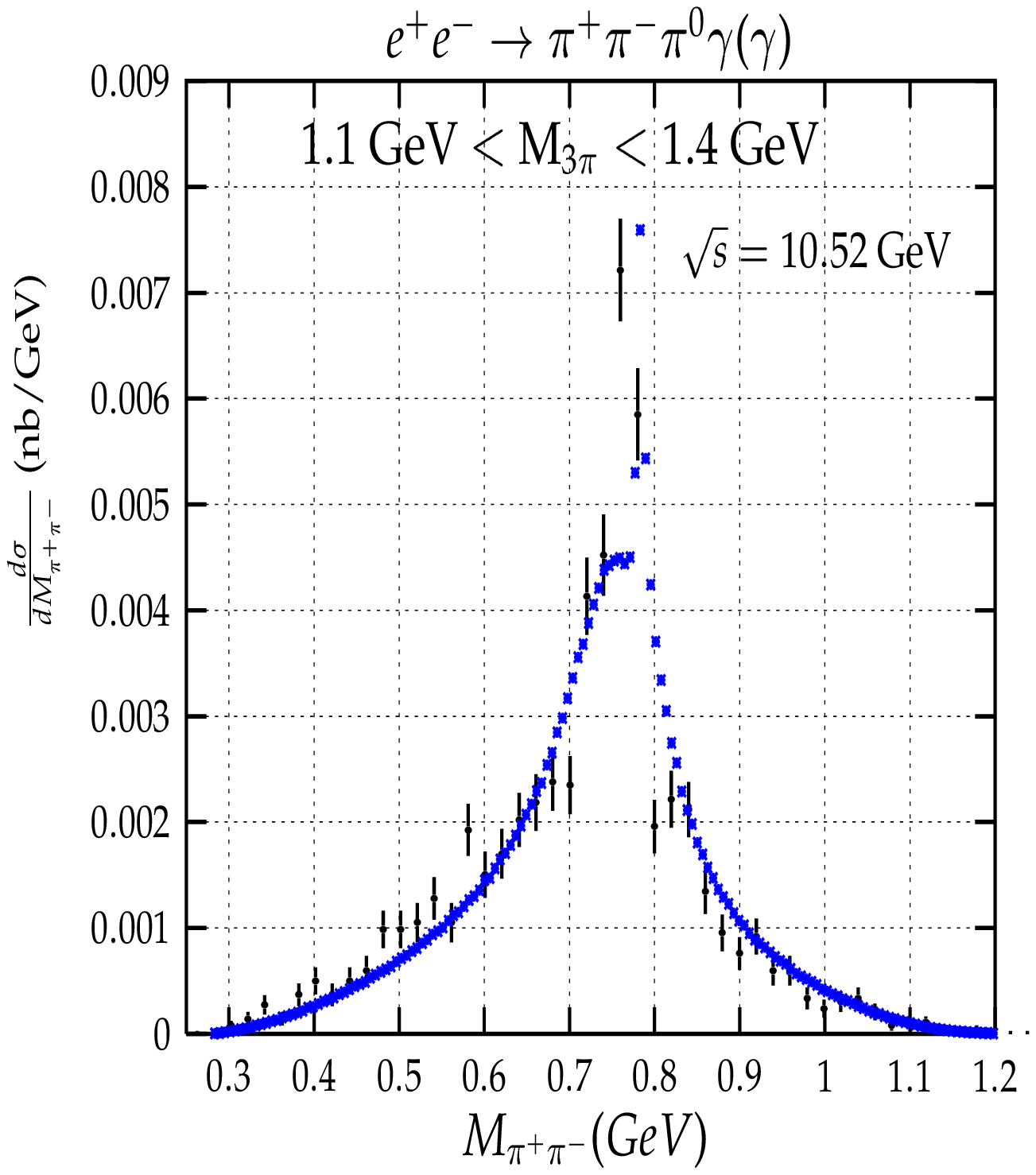,width=6.05cm} 
\epsfig{file=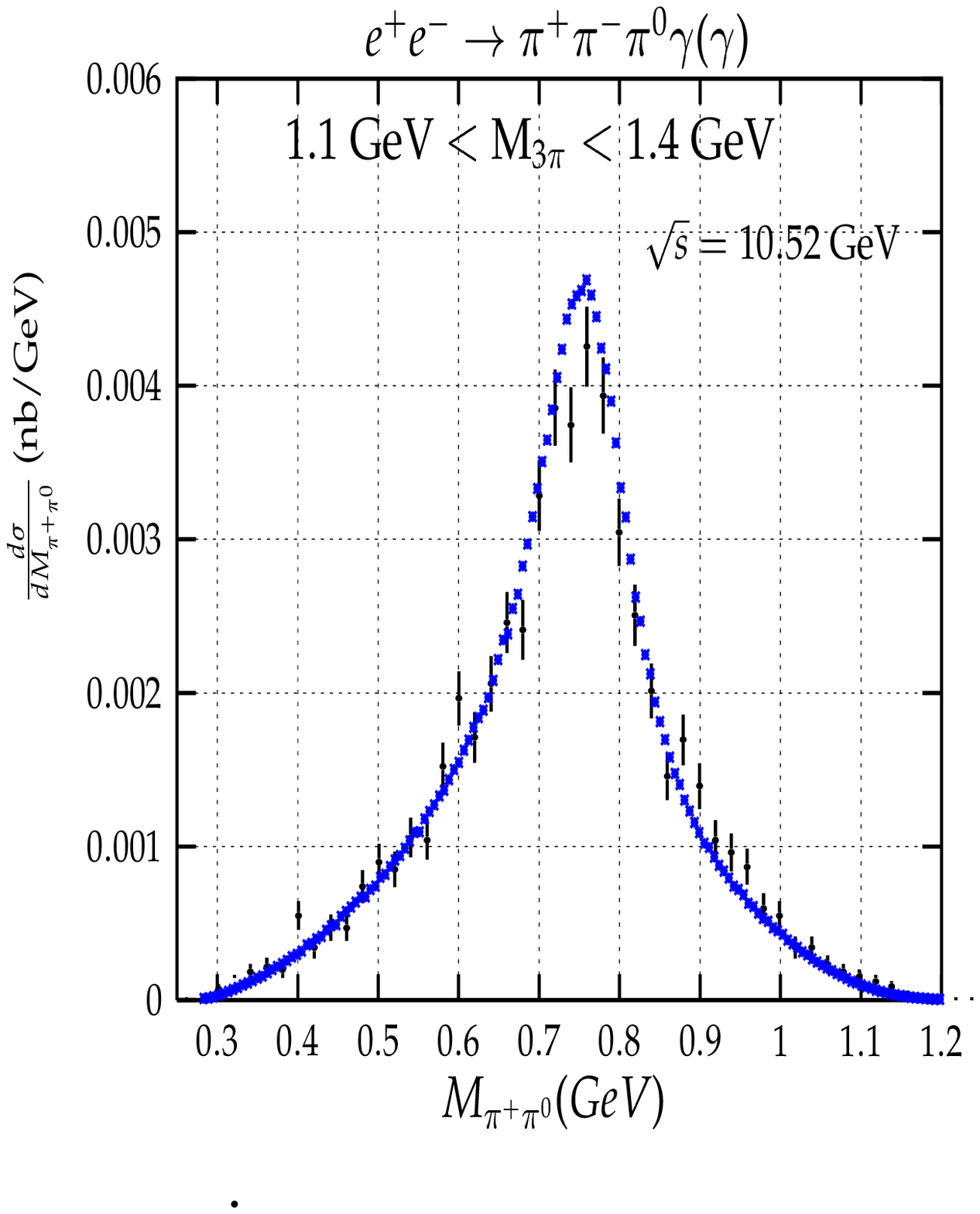,width=5.95cm} 
\caption{ The two-pion invariant mass distributions:
  BaBar data \cite{Aubert:2004kj} (points with error bars) and PHOKHARA5.0
 predictions \cite{new_3pi} (filled circles) .}
\label{BaBar}
\end{center}
\end{figure}

 In the newly released version of 
  the PHOKHARA Monte Carlo generator (5.0) 
  three new hadronic channels were added: $\pi^+\pi^-\pi^0$,
  $K^+K^-$ and $\bar K K$. For the $K^+K^-$ both initial and final
 state photon(s) radiation was taken into account,
 while for the $\pi^+\pi^-\pi^0$
 and the $\bar K K$
  only initial state photon(s)
 emission was considered. The kaon hadronic current
 was adopted from \cite{Bruch:2004py}, while a detailed analysis
 of all existing data on the $e^+e^-\to \pi^+\pi^-\pi^0$ cross section
 was performed in \cite{new_3pi}. 
 The constructed  model allows not only for an excellent
 fit to the cross section and a good description of two pion invariant
 mass distributions (see Fig.~\ref{BaBar}), but many three-meson couplings
 were extracted separately from that fit making possible predictions
 of various decay rates and cross sections \cite{new_3pi}.

\subsection{Next to leading order radiative corrections to 
 the reaction $e^+e^-\to \mu^+\mu^-\gamma$}
\begin{figure}[ht]
\begin{center}
\epsfig{file=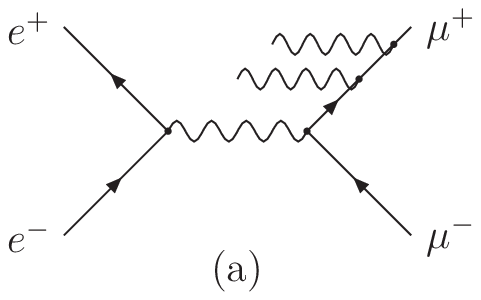,width=4cm} 
\epsfig{file=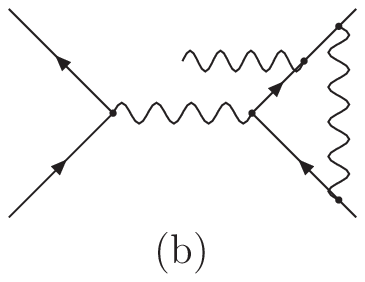,width=4cm} 
\epsfig{file=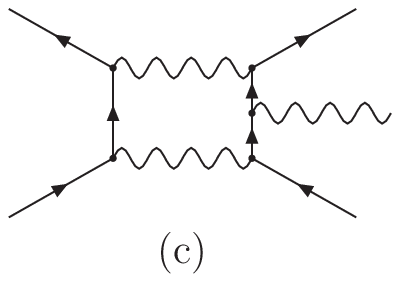,width=4cm} 
\caption{ NLO contributions to the $\sigma(e^+e^- \to \mu^+\mu^-\gamma)$
  missing in the PHOKHARA 5.0 code.}
\label{diag_mu}
\end{center}
\end{figure}

The reaction
\bea 
 e^+(p_1)e^-(p_2) \to \mu^+(q_1)\mu^-(q_2)\gamma(k)
\label{mumu}
\eea
 may serve as a luminosity
 monitoring process for the radiative return method. If this method is used
 one measures 
 the ratio
\bea
{{\cal R}(s)
 =\frac{\sigma(e^+e^-\rightarrow hadrons +\gamma)}
 { \sigma(e^+e^-\rightarrow \mu^+ \mu^- +\gamma)}} \ ,
\label{Rmu}
\eea
 and for the extraction of the  $\sigma(e^+e^-\rightarrow {\rm hadrons})$
 from the data the theoretical knowledge of both
processes is needed. Due to a complicated experimental setup that
 piece of information
has to be provided in a form of event generators and the NLO 
radiative corrections to both processes are indispensable to provide
 accurate theoretical predictions. Already in \cite{PHOKHARA_mu} a part
of the NLO(FSR) radiative corrections was implemented and the missing parts
 of the generator consist of diagrams shown schematically in
 Fig.~\ref{diag_mu}.
All details of the calculations and implementation in the Monte Carlo event
 generator PHOKHARA will be presented in separate publications \cite{new_mu}
 while in this paper
  few details and the present status of this project is sketched.
  The contributions from diagrams in Fig.~\ref{diag_mu}a are implemented
 using helicity amplitude method in an analogous way as it was done
 for two photon ISR contributions \cite{Szopa}. The contributions
 from Fig.~\ref{diag_mu}b, even if in principle are analogous to the
 ISR virtual corrections calculated in \cite{PHradcor}, have to be treated
 in a different way as the generator has to work also for DAPHNE energy
 where the muon mass is not a small parameter. The expansions used
 for the ISR, which shorten the final result enormously, cannot be applied
 and even if the whole result is known further work is
 required to construct formulae for calculation of the radiative corrections
 fast enough for a Monte Carlo event generator. In both cases the
 helicity amplitudes can be written as a combination of 14 terms if
 the gauge invariance and the current conservation is used. The way 
 one writes the result is not unique and we have chosen to write explicitly
 the part proportional to the Born amplitude plus suitably chosen symmetric
 and antisymmetric coefficients. The ISR corrections
 read then
\bea
 {\cal M}_{ISR} \sim \bar u(q_2)\gamma_\mu v(q_1)\cdot
   \sum_{i=1}^{14} F_i \cdot \bar v(p_1)  S_i^\mu u(p_2) \ ,
 \label{ampISR}
\eea
where 
 the coefficients $F_i$ contain loop corrections. The FSR corrections
 have analogous structure.

The functions $S_i^\mu$ read 
\bea
S_1^\mu = \frac{1}{2}
\left(\frac{2p_1\epsilon^*-\ta\epsilon^*\ta k}{p_1k}\gamma^\mu
 -\gamma^\mu\frac{2p_2\epsilon^*-\ta\epsilon^*\ta k}{p_2k}\right)\ , \ 
  S_2^\mu = \ta k\ta\epsilon^*\gamma^\mu\ , 
\nonumber \\
 \ S_3^\mu = \ta k\ta\epsilon^*p_+^\mu\ , \
 \ S_4^\mu = \ta k\ta\epsilon^*p_-^\mu\ , \ 
 \ S_5^\mu = \ta\epsilon^*p_+^\mu-\ta k\epsilon^{*\mu}\ , \ 
\nonumber \\
 \ S_6^\mu = \left[ \ta k \ p_+\epsilon^* -
 \ta\epsilon^*  p_+k\right]\gamma^\mu\ ,
 \ S_7^\mu = \left[ \ta k \ p_-\epsilon^* -
 \ta\epsilon^* p_-k\right]\gamma^\mu\ ,
\nonumber \\
 \ S_8^\mu =  \ta k 
  \left[ p_1\epsilon^*\  p_2k - p_2\epsilon^*\  p_1k\right]p_-^\mu\ , \
 \ S_9^\mu =  \ta k 
  \left[ p_1\epsilon^*\  p_2k - p_2\epsilon^*\  p_1k\right]p_+^\mu\ , \
\nonumber \\
 \ S_{10}^\mu = \ta\epsilon^* p_+^\mu - \ta k 
    \left[ \frac{p_1\epsilon^*}{p_1k}p_2^\mu
    + \frac{p_2\epsilon^*}{p_2k}p_1^\mu\right]\ , \
 \ S_{11}^\mu = \ta\epsilon^* p_-^\mu - \ta k 
    \left[ \frac{p_1\epsilon^*}{p_1k}p_2^\mu
    - \frac{p_2\epsilon^*}{p_2k}p_1^\mu\right]\ , \
\nonumber \\
 \ S_{12}^\mu =  \epsilon^{*\mu}- \frac{p_1\epsilon^*}{p_1k}p_2^\mu
    - \frac{p_2\epsilon^*}{p_2k}p_1^\mu\ , \
\nonumber \\
 \ S_{13}^\mu =   
  \left[ p_1\epsilon^*\  p_2k - p_2\epsilon^*\  p_1k\right]p_-^\mu\ , \
 \ S_{14}^\mu =  
  \left[ p_1\epsilon^*\  p_2k - p_2\epsilon^*\  p_1k\right]p_+^\mu\ , \
\nonumber
\eea
where $p_\pm = p_1\pm p_2$ and $\epsilon$ is the photon polarisation vector.

 Within that 
parameterisation all the coefficients, but the $F_1$,
 are free from ultraviolet and infrared singularities, as the $S_1$
has exactly the spinor structure of the Born (ISR) amplitude. Moreover
 most of the coefficients vanish in the massless limit and they are numerically
 important only for the configurations with the photon collinear to one of the
 initial leptons.

 The diagrams in  Fig.~\ref{diag_mu}c consist of box and pentabox
 diagrams. The pentabox tensor integrals (up to the third rank), which
 has appeared in the calculations, were reduced in D-dimension
 to standard  box diagrams (tensor integrals up to rank two)
 with method equivalent to
 the one presented in \cite{DenDit}, even if in this particular case
 the reduction is simple due to the symmetry of the integrals.
 Further reduction
 is done using standard Passarino-Veltman reduction to scalar integrals.
 The reduction of two remaining pentabox scalar integrals to the
 box scalar integrals has introduced simple denominators ($kp_1-kp_2$)
 and ($kq_1-kq_2$), which have zeros in the physical phase space.
 That problem will be solved by means of the expansion of the resulting
 around the mentioned zeros. Again here the remaining problem is
 the size of the result after the tensor reduction and further
 work is required to produce formulae, which can be used within a Monte
 Carlo program.  

\section{Summary}

 A short review of experimental results obtained by means of the radiative
 return method is presented, with an extensive discussion of
 the theoretical basis of the method. The new version of the PHOKHARA
 Monte Carlo event generator (PHOKHARA5.0) is presented. The status
 of the work on the NLO radiative corrections to the reaction
  $e^+e^- \to \mu^+\mu^- \gamma$, a luminosity monitoring
 process for the radiative return method, is also outlined. 

\vskip 0.1 cm

{\bf Acknowledgements}

 The publication is based in a big part on results obtained in collaboration 
 with
 J.~H.~K\"uhn and G.~Rodrigo. The authors are grateful for many
 useful discussions concerning experimental aspects of the radiative
 return method to members of the KLOE collaboration, mainly 
 Cesare Bini, Achim Denig, Wolfgang  Kluge, Debora Leone, Stefan Miller,
 Federico Nguyen and Graziano Venanzoni.

%*******************************************************************

\end{document}